\documentclass[10pt,a4paper,twocolumn,english,prb,aps,showpacs,floatfix,groupedaddress,superscriptaddress]{revtex4-1}
\usepackage{graphicx}
\usepackage{epsfig}
\usepackage[english]{babel}
\usepackage{amsmath}
\usepackage{amssymb}
\usepackage{amsfonts}
\usepackage{longtable}
\usepackage{dcolumn}% Align table columns on decimal point
\usepackage{bm}
\usepackage{bbm}
\usepackage{nicefrac}
\usepackage{color,array}
\usepackage{colortbl}
\usepackage{dsfont}
\usepackage{mathtools}
\usepackage{subfigure}  % use for side-by-side figures
\usepackage{soul}

\usepackage{xcolor,cancel}

\usepackage[breaklinks,colorlinks=true,citecolor=blue,linkcolor=blue]{hyperref}
\def\dag{{\dagger}}
\def\vpdag{{\vphantom{\dagger}}}

\begin{document}

\newcommand{\be}{\begin{equation}}
\newcommand{\ee}{\end{equation}}
\newcommand{\bearr}{\begin{eqnarray}}
\newcommand{\eearr}{\end{eqnarray}}
\newcommand{\bseq}{\begin{subequations}}
\newcommand{\eseq}{\end{subequations}}
\newcommand{\nn}{\nonumber}
\newcommand{\reqn}{\eqref}

\title{Singlet exciton condensation and bond-order-wave phase\\ in the extended Hubbard model}

\author{Mohsen Hafez-Torbati}
\email{mohsen.hafez@tu-dortmund.de}
\altaffiliation{Current address: Institut f\"ur Theoretische Physik, 
Goethe-Universit\"at, 60438 Frankfurt/Main, Germany}

\author{G\"otz S. Uhrig}
\email{goetz.uhrig@tu-dortmund.de}
\affiliation{Lehrstuhl f\"ur Theoretische Physik I, 
Technische Universit\"at Dortmund,
Otto-Hahn-Stra\ss e 4, 44221 Dortmund, Germany}

\date{\rm\today}

\begin{abstract}
The competition of interactions implies the compensation of standard mechanisms
which leads to the emergence of exotic phases between conventional phases.
The extended Hubbard model (EHM) is a fundamental example for the 
competition of the local Hubbard interaction and the nearest-neighbor density-density interaction, 
which at half-filling and in one dimension leads to a bond order wave (BOW) between 
a charge density wave (CDW) and a quasi-long-range order Mott insulator (MI).
We study the full momentum-resolved excitation spectrum of the one dimensional EHM in the CDW phase 
and clarify the relation between different elementary energy gaps.
We show that the CDW-to-BOW transition is driven by the 
softening of a singlet exciton at momentum $\pi$. The BOW is realized 
as the condensate of this singlet exciton. 
\end{abstract}

% 71.10.Fd 	Lattice fermion models (Hubbard model, etc.)
% 71.10.Li 	Excited states and pairing interactions in model systems
% 71.30.+h 	Metal-insulator transitions and other electronic transitions
% 74.20.Fg 	BCS theory and its development

\pacs{71.30.+h,71.10.Li,71.10.Fd,74.20.Fg}

\maketitle

\section{Introduction}

Strong interactions among electrons can lead to the emergence of collective phenomena 
such as the stabilization of new phases of matter which host non-trivial elementary 
excitations \cite{Balents2010,Senthil2004}.
The role of onsite Hubbard interaction and its competition with different kinetic terms 
is widely investigated \cite{Varney2009,Meng2010,Sorella2012,Yang2010,
Cocks2012,Ebrahimkhas2015,Hafez-Torbati2016}.
However, relatively less attention is paid to the effect of non-local short-range 
interactions such as first or second neighbor density-density interaction terms. 
The possible spontaneous emergence of 
quantum anomalous Hall state (for spinless case) and quantum spin Hall state (for spinfull case) 
on the honeycomb lattice due to first and the second neighbor interactions serves as 
an interesting controversial example in the field of topological Mott insulator  
\cite{Raghu2008,Motruk2015,Scherer2015}.

In order to study quantum phase transitions and to search for non-trivial 
quantum states  it is well-established to analyze effective models with 
competing interaction terms. Their competition compensates the driving
mechanisms of rather trivial phases so that the non-compensated higher order terms 
dominate the physics \cite{Fabrizio1999,Capriotti2001,Sandvik2007}. 
An example is the extended Hubbard model (EHM) at half filling
where two interactions,
namely the onsite Hubbard repulsion $U$ and the nearest-neighbor (NN) repulsion $V$
compete. 

We study the half-filled EHM in one dimension at zero temperature; its Hamiltonian reads 
\begin{eqnarray}
H=&&\!\!\!\! t\sum_{i\sigma} (c^\dagger_{i,\sigma}c^{\phantom{\dagger}}_{i+1,\sigma} + 
{\rm h.c.})
+V\sum_{i}\left( n_{i}-1 \right) \left(n_{i+1}-1 \right)
\nn \\
&+&U\sum_{i} \left( n_{i,\uparrow}-\nicefrac{1}{2} \right) 
\left(n_{i,\downarrow}-\nicefrac{1}{2} \right)
\label{eq:EHM}
\end{eqnarray}
where $c_{i,\sigma}^\vpdag$ and $c^\dagger_{i,\sigma}$ are electron annihilation and creation
operators at site $i$ with spin $\sigma$, respectively. The density operator  
$n_{i,\sigma} := c^\dagger_{i,\sigma}c^{\phantom{\dagger}}_{i,\sigma}$ counts the number of electrons with spin $\sigma$ at site $i$ and 
$n_{i}^\vpdag:=n_{i,\uparrow}\vpdag+n_{i,\downarrow}^\vpdag$.
In the two-fold degenerate CDW regime ($V\gg U$), the gaps to both singlet and triplet excitations are finite. 
In the MI ($U \gg V$), the charge degrees of freedom are frozen and the low-energy physics 
is captured by the Heisenberg model with quasi long-range magnetic order and 
gapless spin excitations.

\begin{figure}[t]
  \centering
  \includegraphics[width=0.95\columnwidth,angle=0]{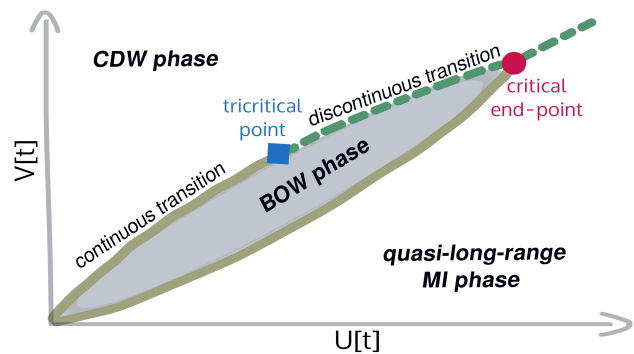}
  \caption{(Color online) Schematic phase diagram of the extended Hubbard model \reqn{eq:EHM} 
  found by bosonization \cite{Nakamura1999,Nakamura2000,Tsuchiizu2002,Tsuchiizu2004,Tam2006}, 
  quantum Monte Carlo \cite{Hirsch1984,Sengupta2002,Sandvik2004}, and density matrix renormalization 
  group method \cite{Jeckelmann2002,Zhang2004,Ejima2007}.
  }
  \label{fig:phase_diagram}
\end{figure}

The phase diagram of the EHM \reqn{eq:EHM}, see Fig.\ \ref{fig:phase_diagram}, 
has been studied extensively using 
bosonization \cite{Nakamura1999,Nakamura2000}, renormalization group 
\cite{Tsuchiizu2002,Tsuchiizu2004,Tam2006}, quantum Monte Carlo (QMC) 
\cite{Hirsch1984,Sengupta2002,Sandvik2004}, 
and density matrix renormalization group (DMRG) 
\cite{Jeckelmann2002,Zhang2004,Ejima2007}. The CDW and the MI are separated 
by the intermediate BOW phase for small to intermediate values of $U$ and $V$. 
For large $U$ and $V$ values beyond a critical end point, 
the BOW disappears and a direct first order transition from the CDW  
to the MI is observed. 
The CDW-to-BOW transition changes from second order 
to first order beyond a tricritical point while the BOW-to-MI transition 
remains second order. Modified models with similar tricritical points have also been 
studied \cite{Lange2015,Ejima2016}. 

The phase transitions in the EHM \reqn{eq:EHM} are determined by computing 
various correlation functions as well as charge and spin gaps. It is reported that at the   
second order CDW-to-BOW transition the charge gap vanishes while the spin gap remains 
finite \cite{Ejima2007}. 
In the MI, the spin gap is zero and holon-antiholon form bound states \cite{Essler2001}. 
In previous investigations, however, the possible formation of an electron-hole bound state with $S=0$, i.e., 
of a singlet exciton, has not been considered. 

In this paper, we present the full momentum-resolved low-energy spectrum of the 
EHM \reqn{eq:EHM} in the CDW phase close to the transition using continuous unitary transformations (CUTs) \cite{Wegner1994,Knetter2000,Kehrein2006}. A rich excitation spectrum comprising two singlet and two triplet bound states is identified. In contrast to the DMRG analysis \cite{Ejima2007}, we find that the second order transition 
from the CDW to the BOW is induced by the vanishing of the optical gap, i.e., the energy of a singlet exciton at total 
momentum $k=\pi$ vanishes. The bond order reflects the condensate of these singlet excitons
and can be understood by a BCS-like mean-field theory.
The spin gap remains finite and is smaller than the charge gap at the transition point.

\section{Charge Density Wave}

The ground state has total spin $S=0$. In order to track phase transitions we
consider four excitation gaps: the 1-particle gap $\Delta_{1}$, the charge gap $\Delta_c$, 
the singlet exciton gap (or 
optical gap) $\Delta_e$, and the spin gap $\Delta_s$ defined by \cite{footnote}
\bseq
\label{eq:gap}
\begin{align}
\label{eq:1particle_gap}
\Delta_{1} &:=E_0^{L+1}+E_0^{L-1}-2E_0^{L}=2(E_0^{L+1}-E_0^{L}) \\
\label{eq:charge_gap}
\Delta_c &:=\frac{1}{2}\left(E_0^{L+2}+E_0^{L-2}-2E_0^{L}\right)=E_0^{L+2}-E_0^{L} \\
\label{eq:singlet_gap}
\Delta_e &:=E_{1,S=0}^{L}-E_0^{L} \\
\label{eq:spin_gap}
\Delta_s &:=E_{1,S=1}^L-E_0^L
\end{align}
\eseq
where $E_0^N$ is the ground state energy at $N$ electrons; at half-filling $N=L$ holds
where $L$ is the number of lattice sites. The energy $E_{1,S}^N$ corresponds to
 the first excited state with total spin $S$ and $N$ electrons. 
The 1-particle gap measures the minimum energy required for adding a \emph{single} electron 
and a \emph{single} hole to the system. The charge gap $\Delta_c$ lies below the 1-particle $\Delta_1$ gap only 
in the case of Cooper-pair formation otherwise they are equal. 
The second equalities in \reqn{eq:1particle_gap} 
and \reqn{eq:charge_gap}  hold due to particle-hole symmetry. 
An electron-hole \emph{pair}  can form a bound state (exciton)
in the singlet and/or in the triplet channel. Its energy defines the singlet 
and the spin gap, respectively. 
We stress that this consideration implies that the singlet and the spin gap
must be equal or smaller than the 1-particle gap. If the gaps are smaller the difference 
in energy is the excitonic binding energy. 
We notice that a charge gap smaller than the spin gap as suggested in DMRG 
analysis \cite{Ejima2007} can only be understood based on electron-electron (hole-hole) bound states.

Different definitions are used for the charge gap in different contexts 
and we have to clarify this point before proceeding.
The singlet exciton gap \reqn{eq:singlet_gap} and the spin gap \reqn{eq:spin_gap} 
can be extracted from the Fourier transform of the charge-charge $\langle n_{i}n_{i+d}\rangle$ 
and the spin-spin $\langle S^z_{i}S^z_{i+d}\rangle$ correlation functions, respectively, as calculated for 
the 1D EHM by QMC method in Refs. \onlinecite{Sengupta2002,Sandvik2004}. 
What is called ``charge gap'' in these references is equivalent to 
our singlet exciton gap Eq. \reqn{eq:singlet_gap}. 
The singlet exciton gap and the spin gap are also the gaps addressed in bosonization 
\cite{Nakamura1999,Nakamura2000,Tsuchiizu2002,Tsuchiizu2004} as the bosonized field always create {\it pair} 
of electron and hole. We notice that for the proper treatment of the 1-particle gap \reqn{eq:1particle_gap} 
and the charge gap \reqn{eq:charge_gap} in the bosonization approach the explicit consideration of Klein factors 
would be necessary \cite{Delft1998}.

\begin{figure*}[t]
  \centering
  \includegraphics[width=2.03\columnwidth,angle=0]{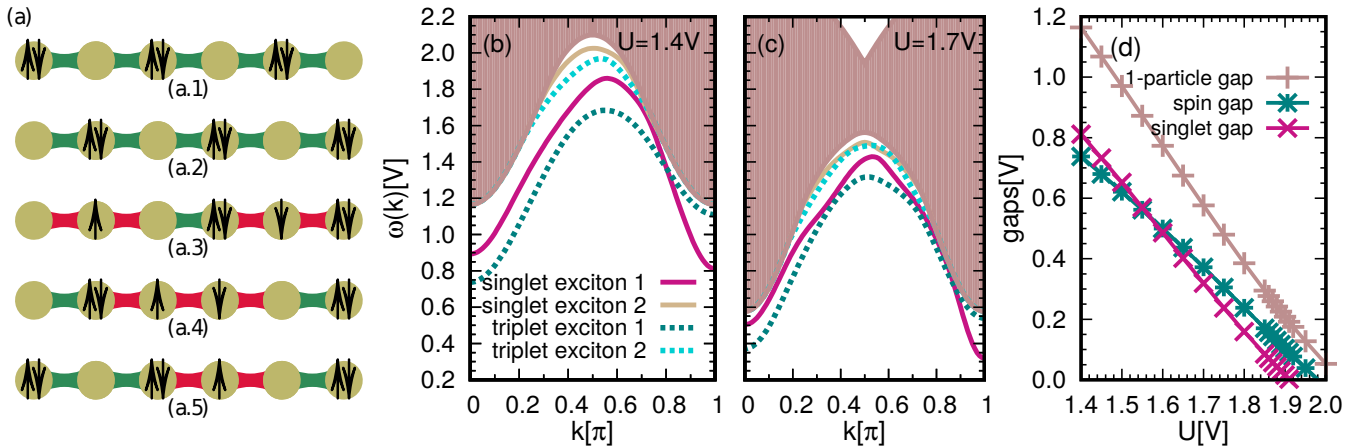}
  \caption{(Color online) (a) Schematic representation of the two degenerate CDWs (a.1 and a.2),
	of an excited  electron-hole pair (a.3 and a.4), and of a single domain-wall (a.5) 
	on a piece of chain of six sites. 
	The bonds low in energy due to $V$ in \eqref{eq:EHM} 
	are shown in green, the ones high in energy
	are shown in red. Clearly, configurations where electron and hole are close
	together are favored suggesting binding.
  Excitation spectrum of the EHM \reqn{eq:EHM} 
in the CDW for $U=1.4V$ (b) and $U=1.7V$ (c). 
(d) The 1-particle gap $\Delta_1$, singlet gap $\Delta_e$, and spin gap $\Delta_s$,
defined in  \eqref{eq:gap},  vs.\  the onsite interaction $U$. The hopping $t$ 
is set to $0.5V$ and the order of the deepCUT is 10.}
  \label{fig:cdw_full}
\end{figure*}

In the atomic limit ($t=0$) and for $2V>U$ the ground state is a two-fold 
degenerate CDW where empty and fully occupied sites alternate,
see Figs.\ \ref{fig:cdw_full}a.1 and \ref{fig:cdw_full}a.2. The system 
becomes excited if an electron hops from an occupied site to an empty one creating 
an electron-hole pair in Figs.\ \ref{fig:cdw_full}a.3 and \ref{fig:cdw_full}a.4. 
The 1-particle gap is given by $\Delta_{1}^\vpdag=4V-U$ as can be read off
from Fig.\ \ref{fig:cdw_full}a.3 where electron and hole are separated.
To minimize its energy, the electron-hole pair can form a bound state on NN sites so
that  the singlet (and the spin gap) is given by $\Delta_s^\vpdag=\Delta_e^\vpdag=3V-U$,
see Fig.\ \ref{fig:cdw_full}a.4. A single domain-wall separating the two degenerate ground 
states is depicted in Fig.\ \ref{fig:cdw_full}a.5, requiring the excitation energy $2V-U/2$, 
i.e., $\Delta_1/2$. 

From this simple argument, one can deduce that the NN interaction strongly favors the formation 
of neutral exciton. 
The degeneracy of the singlet and the triplet gap in the atomic
limit is lifted due to NN hopping. We show that these bound states survive even 
close to the CDW-to-BOW transition. A similar scenario of exciton formation due to NN interaction 
has been found in related models \cite{Hafez2010b,Hafez2011}.

We take the CDW in Fig.\ \ref{fig:cdw_full}a.1 as reference state. 
The electron-hole transformation 
$T^{\rm (e-h)}:c_{i,\sigma}^{\dag} \rightarrow h_{i,\sigma}^\vpdag $ on the odd 
sublattice  expresses the EHM \reqn{eq:EHM} in terms of quasi-particles (QPs). 
This means that any creation operator after the transformation stands for the creation of
an excitation: adding an electron to an empty site or adding a hole to
an fully occupied site. Then, the electron and hole 
operators are uniformly denoted by the fermion operator $f_{i,\sigma}^{(\dag)}$. 
After the local transformation 
$T^{\rm (l)}:f_{j,\sigma}^{\dag} \rightarrow e^{i\frac{\pi}{2}j}
e^{-i\frac{\pi}{4}}f_{j,\sigma}^{\dag}$ 
the EHM can be written as 
\bearr
H = &&\!\!\!\!\!\frac{U-4V}{4} \sum_i \mathds{1} 
+ \frac{4V - U}{2} \sum_{i,\sigma} f^\dagger_{i,\sigma} f_{i,\sigma}^{\phantom{\dagger}} 
\nn \\
&+& U \sum_i f^\dagger_{i,\uparrow}f^\vpdag_{i,\uparrow}f^\dagger_{i,\downarrow}
f^\vpdag_{i,\downarrow} \!- \!V \sum_{i\sigma\beta} f^\dagger_{i,\sigma}f^\vpdag_{i,\sigma}
f_{i+1,\beta}^\dag f_{i+1,\beta}^\vpdag  
\nn \\
&+& t \sum_{i,\sigma} ( f^\dagger_{i,\sigma} f^\dagger_{i+1,\sigma} + {\rm h.c.} ) .
\label{eq:EHM_qp}
\eearr
The local transformation $T^{\rm (l)}$ has restored the full translational symmetry
facilitating the subsequent analysis.
In the QP representation, the original hopping term has become a Bogoliubov term  creating 
a singlet pair of fermions on NN sites. 
We stress that the NN electron-electron interaction in \eqref{eq:EHM_qp}
has acquired a minus sign indicating attraction between the original electron and hole.

To eliminate the Bogoliubov terms which change the number of
QPs we employ the directly evaluated enhanced perturbative CUT (deepCUT) \cite{Krull2012}.
The resulting effective model allows us to analyze the complete momentum-resolved 
excitation spectrum of the Hamiltonian \reqn{eq:EHM_qp}. 
The CUT is performed in the thermodynamic limit and is known as a powerful approach to compute 
excitation spectra and spectral densities 
\cite{Knetter2001,Knetter2004,Hafez-Torbati2014,Powalski2015}.
We treat the Bogoliubov term as the perturbation in the deepCUT formalism \cite{Krull2012} 
so that the flow equations are truncated in powers of the hopping $t$. The same  
symmetries and simplification rules can be used as in Ref.\ \onlinecite{Hafez-Torbati2014}.

In the CUT method, the Hamiltonian is mapped to an effective one by a unitary transformation 
which depends on an auxiliary parameter $\ell$. The transformed Hamiltonian satisfies
the flow equation \cite{Wegner1994,Kehrein2006,Knetter2000}
\be
\partial_\ell H(\ell)=\left[\eta(\ell), H(\ell) \right],
\ee
where the antihermitian operator $\eta(\ell)$ is the generator of the flow and  
determines the essence of the transformation. 
We decompose the Hamiltonian into different 
parts which create and annihilate specific numbers of QPs \cite{Knetter2000,Krull2012}:
\be 
H(\ell)=\sum_{n,m} H_{n:m}(\ell)
\ee
where $H_{n:m}$ creates $n$ and annihilates $m$ QPs. The reduced generator \cite{Fischer2010} 
\be 
\eta_{p:x}^{\vphantom{\dagger}} = \sum_{m=0}^{x} \sum_{n>m} \left( H_{n:m} - {\rm h.c.} \right)
\ee
allows us to decouple the first $x$ QP sectors from higher sectors.

Using the reduced generator $\eta_{p:2}^{\vphantom{\dagger}}$ to decouple up to two QPs sector 
in the EHM leads to a diverging flow
because the decoupling of the subspaces with two QPs is 
difficult if binding phenomena prevail. Hence we used the  $\eta_{p:1}^{\vphantom{\dagger}}$ generator 
instead and implemented the diagonalization in the 
$2$-QP subspace \cite{Hafez-Torbati2014}. This means that the
off-diagonal terms linking the $2$-QP sector to $4$ and higher QP sectors are neglected. 
This procedure can be understood as 
a variational approximation for the {\it effective Hamiltonian} derived from deepCUT. The neglected terms 
would only increase the binding energies, thereby enhancing the effects discussed
in this work. Moreover, we know from data in order $6$ close to the transition
where the $\eta_{p:2}^{\vphantom{\dagger}}$ generator still converges, that the obtained results are 
quantitatively close
to the ones obtained using the $\eta_{p:1}^{\vphantom{\dagger}}$ generator so that we conclude that the
neglected off-diagonal elements are of minor importance.

In Figs.\ \ref{fig:cdw_full}b ($U=1.4V$) and \ref{fig:cdw_full}c ($U=1.7V$)
the excitation spectrum of the EHM \reqn{eq:EHM} is depicted in the CDW phase.
The neutral singlet (triplet) excitons are specified by solid (dashed) lines. 
The solid areas indicate the electron-hole continua constructed from 
the single fermion dispersion. 
A rich excitation spectrum comprising two singlet and two triplet neutral excitons is identified. 
We have not found any electron-electron (hole-hole) bound state in the entire Brillouin zone.
The exciton 1 exists almost in the whole Brillouin zone 
while the exciton 2 is present only close to $k=\pi/2$ (lattice constant
is set to unity).
We ascribe the small wiggles close to $k={\pi/2}$ to the truncation in finite order. 
For $U=1.4V$,  the singlet exciton 1 takes its minimum energy 
at $k=\pi$. This minimum is higher in energy than the minimum 
of the triplet exciton 1 at $k=0$.  Increasing, however, the 
Hubbard interaction to $U=1.7V$ one discerns in Fig.\ \ref{fig:cdw_full}c that the 
lowest excited state is the singlet exciton 1 at $k=\pi$. 
It is this singlet exciton which becomes softs at the transition to the BOW
upon increasing $U$ further. Beyond the transition it forms a macroscopic
condensate, i.e., the BOW. The same behavior is found in order $6$ and $8$. 
This is consistent with bosonization \cite{Nakamura1999,Nakamura2000,Tsuchiizu2002,Tsuchiizu2004} 
and QMC analysis \cite{Sengupta2002,Sandvik2004} which suggest vanishing of 
a {\it neutral spinless} gap at the CDW-to-BOW transition, but disagrees with 
DMRG \cite{Ejima2007} which proposes the vanishing of the charge gap \reqn{eq:charge_gap}.

According to the definitions in \reqn{eq:gap}, the 1-particle gap is given by 
the lowest energy of the electron-hole continuum which occurs at $k=0$ and $k=\pi$. 
The charge gap equals the 1-particle gap as no electron-electron bound state is found.
The singlet exciton 1 at $k=\pi$ and the triplet exciton 1 at $k=0$ define the 
singlet and the spin gap, respectively. 
This clarifies the difference between the 1-particle gap $\Delta_1$ and 
the singlet gap $\Delta_e$. The dependences of the gaps on $U$ is 
presented in Fig.\ \ref{fig:cdw_full}d. For $U\lesssim 1.57V$ 
the lowest excitation has $S=1$ while for $U\gtrsim 1.57V$ it has $S=0$. 
The singlet gap vanishes at the transition $U_{c1}\simeq 1.91V$
while the 1-particle gap remains finite and {\it larger} than the spin gap. 
This modifies the currently used scenario where the charge gap is zero at finite spin gap 
at the CDW-to-BOW transition \cite{Ejima2007}. 
Note that the results in Fig.\ \ref{fig:cdw_full}d are valid only up to the 
transition.

\section{Singlet Exciton Condensation and Bond Order Wave}
% \label{sec:mft}

Once the energy of an exciton falls below zero its creation lowers the 
total energy of the system. Hence, more and more of them will be created
leading to a macroscopic occupation: a condensate is formed. This
continued exciton creation comes to an end due to residual repulsive interactions
between them. Such interactions exist because only the composite object,
the exciton, behaves like a boson. The internal fermionic structure prevents
two excitons to come too close. 

This physics is captured by a BCS-type mean-field theory applied to the effective Hamiltonian systematically derived by deepCUT beyond the CDW-to-BOW transition at $U=U_{c1}$.
Here, we show in this way that the condensation of the singlet exciton at 
$k=\pi$ leads to the BOW. Of course, the critical fluctuations of the transition
and thus its critical exponents are not accounted for by the BCS theory, but
our focus is here on the driving mechanisms resulting from the fundamental energies
in the system. Critical behavior may be captured by bosonization 
\cite{Nakamura1999,Nakamura2000}, analytical \cite{Tsuchiizu2002,Tam2006} or numerical 
\cite{Jeckelmann2002,Zhang2004,Ejima2007}
renormalization approaches or quantum Monte Carlo \cite{Sengupta2002,Sandvik2004}.
 
To describe the BOW, we consider the effective Hamiltonian from the deepCUT 
up to quartic level
\be 
\label{eq:EHM_eff}
H_{\rm eff}=E_0+\sum_{ij}\Gamma_{j;i}f_{j}^\dag f_{i}^\vpdag
+\sum_{klij}\Gamma_{kl;ij}f_{l}^\dag f_{k}^\dag f_{i}^\vpdag f_{j}^\vpdag,
\ee
where the range of hopping and interaction processes in \reqn{eq:EHM_eff} is limited 
by the order of the truncation. The quartic Hamiltonian \reqn{eq:EHM_eff} captures the condensation of 2-QP bound states.  
In the BCS analysis, we allow for finite expectation values 
$\langle f_{i,\sigma}^\dag f_{i+m,\sigma}^\dag \rangle$ and 
$\langle f_{i,\sigma}^\dag f_{i+n,\sigma}^\vpdag \rangle$ where $m$ and $n$ 
are restricted to odd and even numbers, respectively, due to 
the conservation of the total charge. 
We also allow for broken translational symmetry  
$\langle f_{i,\sigma}^\dag f_{i+m,\sigma}^\dag \rangle \neq 
\langle f_{i+1,\sigma}^\dag f_{i+m+1,\sigma}^\dag \rangle$
to account for the possibility of a BOW. \cite{Hafez-Torbati2014}

The bilinear Hamiltonian resulting from the application of 
Wick theorem on \reqn{eq:EHM_eff} reads \cite{Hafez-Torbati2014}
\bearr
\label{eq:BCS}
H&=&{\tilde{\rm E}_0}+ 
\sum_{r\sigma}\sum_{m} \Delta_{m}^r ( :\!f_{r,\sigma}^\dagger f_{r+m,\sigma}^{\dagger}\!: +{\rm h.c.})
\nn \\ 
&+& \sum_{r\sigma}\left(  
t_0^{\vphantom{\dagger}} :\!f_{r,\sigma}^\dagger f_{r,\sigma}^{\vphantom{\dagger}}\!:
\!+\!\sum_{n} t_n^{\vphantom{\dagger}} ( :\!f_{r,\sigma}^\dagger f_{r+n,\sigma}^{\vphantom{\dagger}}\!:+{\rm h.c.})
\right)
\eearr
where the Bogoliubov prefactor $\Delta_{m}^r$ changes from odd to even sublattice. 
We consider $\Delta_{m}^r=\Delta_{m}^A$ for $r$ even and $\Delta_{m}^r=\Delta_{m}^B$ for $r$ odd.
The prefactors $\tilde{\rm E}_0$, $t_{n}^{\vphantom{\dagger}}$, $\Delta_{m}^A$, and $\Delta_{m}^B$ 
depend on the coefficients of the effective Hamiltonian \reqn{eq:EHM_eff} and the bilinear 
expectation values which are to be determined self-consistently.
The BCS Hamiltonian \reqn{eq:BCS} is diagonalized in momentum space by 
a Bogoliubov transformation. After some standard calculations one obtains the 
self-consistent equations
\bseq 
\begin{align}
 \langle f_{r,\sigma}^\dagger f_{r+n,\sigma}^{\vphantom{\dagger}} \rangle &= 
 \frac{1}{\pi} \int_{0}^{\frac{\pi}{2}} \!\!dk ~\frac{\lambda(k)-t(k)}{\lambda(k)} \cos(n k),
 \\
 \langle f_{r,\sigma}^\dagger f_{r+m,\sigma}^{\dagger}\! \rangle &=
 \frac{1}{\pi} \int_{0}^{\frac{\pi}{2}} \!\! \frac{dk}{\lambda(k)}~ 
\Big( 
{\rm Im}(\Delta(k))\sin(m k)
\nn \\
&\hspace{1.3cm} - (-1)^r {\rm Re} (\Delta(k) ) \cos(m k)
\Big).
\end{align}
\eseq
We have defined the functions $t(k)$, $\Delta(k)$, and $\lambda(k)$ as
\bseq 
\begin{align}
t(k)\!&=\!t_0 + 2\sum_{n} t_{n} \cos(n k) \\
\Delta(k)\!&=\!\sum_{m} 
\Big( (\Delta^A_{m}\!-\!\Delta^B_{m})\cos(m k)
\!-\!i (\Delta^A_{m}\!+\!\Delta^B_{m})\sin(m k) \Big) \\
\lambda(k)\!&=\! \sqrt{t^2(k)+|\Delta(k)|^2},
\end{align}
\eseq
where $n$ and $m$ take positive even and positive odd values, respectively.

The BCS analysis is exact in the entire CDW phase 
where the quantum fluctuations are already captured by the deepCUT. 
In the condensate phase, i.e., beyond $U_{c1}$, it is an approximation 
as mentioned above. The energy differences are rendered quite reliably as long
as the system is not shifted too far beyond the transition.

For $U<U_{c1}$, all  expectation values are trivially zero because the deepCUT has 
mapped the ground state of the EHM \reqn{eq:EHM} to the vacuum of QPs. Beyond $U_{c1}$, 
the expectation values become finite. Two degenerate solutions I and II are found corresponding
to the two ways to break the translational symmetry by bond order.
% , ses Figs.\ \ref{fig:scheme}a and  \ref{fig:scheme}b. 
We obtain
$\langle f_{i,\sigma}^\dag f_{i+m,\sigma}^\dag \rangle =
- \langle f_{i+1,\sigma}^\dag f_{i+m+1,\sigma}^\dag \rangle$. The two solutions are related via 
$\langle f_{i,\sigma}^\dag f_{i+m,\sigma}^\dag \rangle_{I}^\vpdag =
- \langle f_{i,\sigma}^\dag f_{i+m,\sigma}^\dag \rangle_{II}^\vpdag$.
The natural order parameter of the BOW is the difference of the expectation values on 
adjacent NN bonds.

\begin{figure}[t]
  \centering
  \includegraphics[width=0.7\columnwidth,angle=-90]{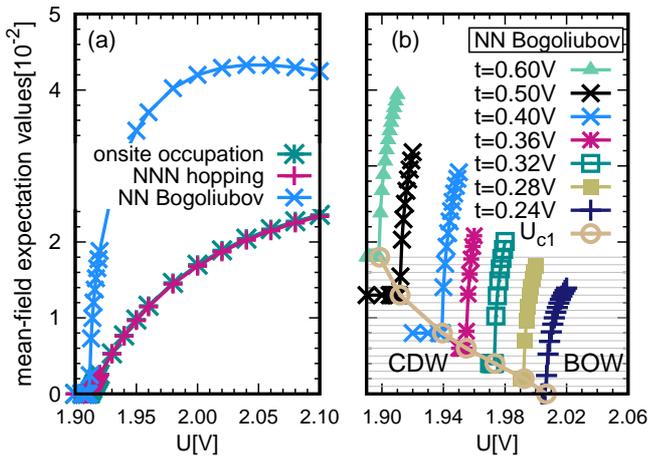}
  \caption{(Color online) (a) The onsite occupation
	$\langle f_{i,\sigma}^\dag f_{i,\sigma}^{\protect\phantom{\dag}} \rangle$, 
the NN Bogoliubov term $\langle f_{i,\sigma}^\dag f_{i+1,\sigma}^{{\dag}} \rangle$, and 
the next-nearest-neighbor (NNN) hopping  
$\langle f_{i,\sigma}^\dag f_{i+2,\sigma}^{\protect\phantom{\dag}} \rangle$ 
vs.\ $U$ for $t=0.5V$. (b) The NN Bogoliubov term 
shifted by $(t-0.24)\times 0.05$ along the $y$-axis for clarity at various  
values of the hopping $t$ as function of $U$. The order of the underlying 
deepCUT is 10.}
  \label{fig:mf}
\end{figure}

In Fig.\ \ref{fig:mf}a we depict the important expectation values 
of the BCS mean-field solution as function of $U$. We stress that the value of $U$ 
where the expectation values become finite matches 
\emph{precisely} the value where $\Delta_e$ hits zero in Fig.\ \ref{fig:cdw_full}d. 
The NN Bogoliubov expectation value displays a square root behavior as usual in mean-field.

In Fig.\ \ref{fig:mf}b, the NN Bogoliubov term is plotted  for various hopping parameters 
versus the Hubbard interaction $U$. From this figure, one can read off the transition line 
CDW-to-BOW phases, $U_{c1}(t)$. We expect the CDW-to-BOW transition to change from 
second order to first order below $t\simeq 0.32V$ based on previous results 
\cite{Sandvik2004,Ejima2007}. From Fig.\ \ref{fig:mf}b, we find a second order transition at least down to $t=0.24V$. Below $t=0.24V$ even the reduced generator $n:1$ diverges. The fact that we can not identify the tricritical point where the character of
the transition changes may either result from the truncation of the effective 
Hamiltonian  \reqn{eq:EHM_eff} to quartic terms 
or from the approximate treatment on mean-field level. 
Recall that finding first order transitions in Landau theory generically requires the 
inclusion of hexatic terms.

Furthermore, it has been proposed by Hirsch that the formation of 
MI ``droplets''  beyond a critical size in the CDW phase 
is responsible for the first order transition in the EHM \cite{Hirsch1984}.
If this is the mechanism of the first order transition one has
to address multi-particle bound states which is beyond the scope
of the present article. The proper description of multi-particle bound
states requires to go beyond quartic level in \reqn{eq:EHM_eff}
because the irreducible interactions of more than two QPs
matter.  

\section{Conclusions and Outlook}

Understanding unusual phases occurring between two more standard phases
is currently a very active topic. A nice example are the phases occurring in
fermionic lattice models such as the ionic Hubbard model or the extended Hubbard
model. In the latter, the two interactions, the onsite one and the nearest-neighbor
one, are competing. Where they compensate in one dimension 
neither the Mott insulator (MI) not the alternating
charge density wave (CDW) occurs, but an alternating bond order wave (BOW). 

In the present work, we have shown that the occurrence of the BOW can
be understood from the CDW as the softening of a singlet exciton at
momentum $k=\pi$. Thus, the bound state of an electron-hole pair 
represents an essential collective mode. Upon passing from the CDW to the BOW,
for instance by increasing $U$, this mode condenses. Since the mode
lives at $\pi$ its condensate naturally displays an alternating order.
It is not accompanied by magnetic order because the condensing mode
does not carry any spin. The same scenario occurred in the ionic Hubbard
model \cite{Hafez-Torbati2014}.

Our finding naturally implies that the singlet exciton gap $\Delta_e$
is \emph{smaller} than the 1-particle gap $\Delta_1$ which reflects the energy needed
to create an electron and a hole excitation independently, i.e., at large
distance. For the spin gap $\Delta_s$ the relation $\Delta_s \le \Delta_1$ 
holds as well because the spin excitation also represents an exciton,
but with $S=1$. The differences $\Delta_1-\Delta_e$ and  $\Delta_1-\Delta_s$ 
are the binding energies of the $S=0$ and the $S=1$ exciton, respectively.

So far, we could not find the first order transition for larger interactions
$U, V$ corresponding to smaller hopping $t$. But we presume that multi-particle
terms need to be included to capture this feature.

A particularly intriguing challenge is to extend the presented analysis
to the two-dimensional extended Hubbard model as the deepCUT method has 
no conceptual problem with dimension and the BCS-mean-field theory is expected 
to work better in higher dimension \cite{Hafez-Torbati2016}. 
There, very little is known about intermediate phases because many theoretical tools
do not work in higher dimensions or only at considerably
larger efforts. But the analogy to the ionic Hubbard model suggests that a rich scenario of 
intermediate phases occurs, breaking first discrete and then continuous
symmetries upon increasing the Hubbard interaction \cite{Hafez-Torbati2016}. 
The possible spontaneous emergence 
of quantum anomalous Hall state and quantum spin Hall state 
on the honeycomb lattice due to competing first and second neighbor interactions is 
another currently controversial issue which calls for future studies. 
\cite{Raghu2008,Daghofer2014,Duric2014,Scherer2015,Motruk2015}

\begin{acknowledgments}
We thank Satoshi Ejima, Fabian Essler, Holger Fehske, and Bruce Normand 
for useful discussions. 
\end{acknowledgments}

\section*{References}
% \bibliographystyle{apsrev4-1}
% \bibliography{references_v7}

\begin{thebibliography}{44}
\bibitem [{foo()}]{footnote}%
  \BibitemOpen
  \href@noop {} {\bibinfo  {journal} {In Ref. 39, we have considered the
  1-particle gap, Eq. (2a), as charge gap. However, we notice that there is no
  difference between the charge gap and the 1-particle gap in the absence of
  Cooper-pair formation}\ }\BibitemShut {NoStop}%
\end{thebibliography}


\end{document}